\documentclass[nonacm]{acmart} % arxiv
\usepackage{balance}
\usepackage{graphicx} % Required for inserting images
\usepackage{array}
\usepackage{makecell}

\AtBeginDocument{%
  \providecommand\BibTeX{{%
    \normalfont B\kern-0.5em{\scshape i\kern-0.25em b}\kern-0.8em\TeX}}}

\usepackage{amsmath}
\usepackage{xcolor}
\usepackage[framemethod=tikz]{mdframed}
\usetikzlibrary{shadows}

\definecolor{mygray}{HTML}{F5F5F5}

\usepackage{booktabs} % For formal tables
\usepackage[flushleft]{threeparttable} % For table notes

\usepackage[utf8]{inputenc}

\begin{document}

\title{Leveraging Large Language Models to Detect Influence Campaigns in Social Media}

% \author{Luca Luceri}
% \authornote{All authors contributed equally to this research.}
% \email{lluceri@isi.edu}
% \affiliation{%
%  \institution{USC Information Sciences Institute}
%   \city{Los Angeles}
%  \state{California}
%  \country{USA}
%   \postcode{90007}
% }
% \author{Eric Boniardi}
% % \authornote{Authors equally contributed to this research.}
% \email{boniardi@isi.edu}
% \affiliation{%
%  \institution{USC Information Sciences Institute}
%   \city{Los Angeles}
%  \state{California}
%  \country{USA}
%   \postcode{90007}
% }

% \author{Emilio Ferrara}
% % \authornote{Authors equally contributed to this research.}
% \email{emiliofe@isi.edu}
% \affiliation{%
%  \institution{University of Southern California \& USC Information Sciences Institute}
%   % \streetaddress{P.O. Box 1212}
%  \city{Los Angeles}
%  \state{California}
%  \country{USA}
%   \postcode{90007}
% }

\author{Luca Luceri}
\authornote{LL, EB, and EF contributed equally to this research.}
\email{lluceri@isi.edu}

\author{Eric Boniardi}
\email{boniardi@isi.edu}

\author{Emilio Ferrara}
\email{emiliofe@usc.edu}

\affiliation{%
 \institution{University of Southern California \& USC Information Sciences Institute}
 \country{Los Angeles, CA, USA}}

\renewcommand{\shortauthors}{Luceri, Boniardi, and Ferrara}

\begin{abstract}

Social media influence campaigns pose significant challenges to public discourse and democracy. Traditional detection methods fall short due to the complexity and dynamic nature of social media. Addressing this, we propose a novel detection method using Large Language Models (LLMs) that incorporates both user metadata and network structures. By converting these elements into a text format, our approach effectively processes multilingual content and adapts to the shifting tactics of malicious campaign actors. We validate our model through rigorous testing on multiple datasets, showcasing its superior performance in identifying influence efforts. This research not only offers a powerful tool for detecting campaigns, but also sets the stage for future enhancements to keep up with the fast-paced evolution of social media-based influence tactics.

\end{abstract}

\maketitle

\section{Introduction}
% The rapid evolution of artificial intelligence, particularly through models like ChatGPT, has introduced new challenges in the digital realm, notably within social media. Key among these is detecting advanced social bots capable of blending with genuine users to manipulate online discourse \cite{ferrara2023social}. These bots, if left unchecked, have a significant impact on public discourse and can shift perceptions on critical issues. The ethical deployment of AI is also a growing concern, as large language models (LLMs), while powerful, carry inherent biases that can inadvertently propagate misinformation or reinforce stereotypes \cite{ferrara2023should}. Furthermore, the advent of Generative AI underscores the dual nature of technology, presenting both extraordinary potential and risks of misuse for malicious purposes, thus amplifying the complexities of digital influence operations \cite{ferrara2023genai}. Navigating this AI-influenced landscape demands a commitment to ethical AI practices, robust safeguards, and ongoing research.

While AI’s potential in diverse fields is unprecedented, its ethical implications, especially in the context of shaping reality, cannot be understated \cite{yang2023anatomy,ezzeddine2022characterizing}. 
For example, within the modern digital ecosystem, malicious actors who aim at reshaping public opinion using influence campaigns have acquired a new powerful weapon: Tools such as ChatGPT can significantly complicate the task of distinguishing between human-generated content or AI content \cite{augenstein2023factuality}. 
Large Language Models (LLMs) can be used to manufacture certain narratives or spread disinformation \cite{ferrara2023should}. 
The use of automation and synthetic personas to steer public opinion on social networks is of particular concern \cite{ferrara2023social}. 
Furthermore, the emergence of Generative AI brings about a new layer of complexity, enabling the creation of potentially highly persuasive messages whose detection is a technical challenge that is almost insurmountable \cite{ferrara2023genai, augenstein2023factuality}.

% \subsection*{Contributions of this work}
This paper addresses the complexities of AI-driven influence campaigns, underscoring the need for diligent research, adherence to ethical AI practices, and the creation of robust countermeasures to maintain the integrity of online dialogue. 
% Specifically, we investigate the use of open-source LLMs in detecting influence campaigns and propose innovative methods to identify such activities across various national contexts.
In particular, we explore the use of open-source LLMs for detecting influence campaigns, introducing novel methods capable of recognizing such campaigns that originate from various nations. Our goal is to surpass the current state-of-the-art in detecting influence campaigns by leveraging content metadata and network information, such as retweet networks and network centralities. To this end, we use a manually labeled dataset to construct supervised models, drawing from a wide spectrum of sources, including Twitter’s Information Operations datasets \cite{gadde2020additional, nwala2022general} and data related to Russian Twitter trolls \cite{addawood2019linguistic}.
Our paper addresses the following research questions (RQs):

\smallskip\noindent\textbf{RQ1:} \textit{To what extent are current state-of-the-art methods effective in detecting information operation campaigns and do these methods experience any degradation in performance when applied to new (unseen) influence campaigns and tactics?}

We find that although state-of-the-art methods are effective, their performance diminishes when they are applied to newer datasets, suggesting that LLMs' potential to adapt may enable them to evade these techniques. This highlights the imperative for innovative dynamic detection methods.

\smallskip\noindent\textbf{RQ2:} \textit{Do techniques based on LLMs offer superior classification performance compared to state-of-the-art methods, and does the integration of various types of information, such as content, network information, and user metadata, enhance classification performance?}

Our findings indicate that LLM-based techniques achieve comparable, and often superior, performance in unseen influence campaigns and tactics. Nonetheless, the performance declines when multiple information types are combined as a single input, underscoring the need for further research to refine the integration of multi-inputs that enhance overall results.

\smallskip\textit{Summary of contributions:} In this article, we conduct an analysis of influence campaigns emanating from four distinct countries. We propose innovative methods for detecting drivers of influence operations and benchmark them against state-of-the-art methods. In doing so, we contribute valuable insights and directions for future research, particularly focusing on the application of open-source LLMs to identify malicious entities within influence campaigns.

\section{Related Work}
The landscape of social media is increasingly shaped by influence operations, often orchestrated by bots and coordinated accounts, representing a key concern in today's digital environment. 
The COVID-19 pandemic underscored the tangible effects of influence operations \cite{nogara2022disinformation}, with Rao et al. \cite{rao2021political} examining the relationship between political biases and antiscience sentiment in online discourse, proposing that influence operations may have contributed to shaping these discussions. Similarly, Jiang et al. \cite{jiang2021social} investigated the amplification of polarization and echo chambers on social media during the pandemic, underscoring the role of influence operations in establishing beliefs and limiting the diversity of information.
The significance of bot-driven influence and the need for sophisticated detection mechanisms were highlighted by Yang et al. \cite{yang2022botometer}, emphasizing the urgency of advanced tools to identify and comprehend bot activities that frequently spearhead these operations \cite{luceri2019evolution}.

Based on the narrative of bot influence, the detection of coordinated accounts, which serve as vectors for influence operations, has emerged as a critical research area \cite{luceri2023unmasking,Pacheco_2020,weber2021amplifying,nizzoli2021coordinated,giglietto2020takes}. Previous work introduced pioneering methodologies to identify such accounts, underlining their importance in concealed influences \cite{sharma2021identifying,luceri2020detecting} and coordinated online campaigns \cite{ezzeddine2022characterizing,suresh2023tracking}. While these approaches have primarily concentrated on identifying coordinated malicious accounts based on behavioral similarities, Addawood et al. \cite{addawood2019linguistic} introduced a content-based approach, which extracts linguistic cues from the messages shared by inauthentic actors to enable their detection.
Complementing this class of approaches, Sapienza et al. \cite{sapienza2018discover} presented DISCOVER, a tool designed to glean insights from online conversations, stressing the need for vigilance against potential malicious operations.

In the realm of employing language models for social media analysis, Kumar et al. \cite{kumar2021content} developed a combined neural network ensemble comprising Text CNN and LSTM models with BERT embeddings to categorize tweets, showcasing the capability of language models in bot detection based on textual content. Advancing the field, Malik et al. \cite{malik2023detect} used BERT, augmented by fine-tuning, to detect propaganda on social media platforms, signifying the effectiveness of semantic and fine-tuned language models.

Acknowledging the vital role of network information in social network analysis and the inherent textual processing nature of language models, it becomes imperative to devise methodologies that can adeptly convert network data into text. Cai et al. \cite{cai2023lmbot} introduced \textit{LM Bot}, illuminating the feasibility of entering network information into a language model for Twitter bot detection. Ye et al. \cite{ye2023natural} provided a formalization to translate graph knowledge into text, demonstrating that natural language processing can efficiently interpret graph data, thus forging new pathways for research in detecting influence campaigns where graph information is crucial.

The advent of new and more potent large language models has carved a niche for their application in the detection of influence campaigns, particularly the open-source variants, which offer the advantage of usage without the need for data sharing, preventing malicious entities from exploiting these models in controlled settings for adversarial learning.

\section{Data}
\subsection{Twitter Information Operations}
In line with recent studies \cite{kong2023interval,nwala2022general,luceri2023unmasking}, this investigation uses data from the Twitter's Information Operations archive \cite{gadde2020additional}. Twitter has released more than 141 datasets detailing information operations, comprising tweets from operators in 21 countries, collected from 2008 to 2021. Operators involved in these campaigns take advantage of a variety of strategies on the platform \cite{luceri2023unmasking}, aimed at different user communities with a wide range of objectives.

These operations are inherently complex, orchestrated by entities ranging from small groups to large collectives, and may involve human agents, automated bots, or compromised accounts. The targets and objectives are equally diverse, encompassing specific groups, entire nations, or broad geopolitical regions, employing tactics from simple spam to sophisticated strategies like coordination, rotation, obfuscation, all adaptable and primarily political in nature. Furthermore, these operations extend beyond Twitter, indicating a widespread challenge in the digital ecosystem.

Researchers at Indiana University have provided access to a control dataset they constructed, which aids in the identification of operators involved in information campaigns \cite{nwala2022general}. They assembled tweets from accounts not directly involved in information operations, but engaging in related discussions concurrently. The hashtags used by the operators were extracted and used to query Twitter’s academic search API, identifying accounts with tweets that match the dates and hashtags of the operators. The timelines for these accounts were then reconstructed, collecting up to 100 tweets from the same dates as the operators. 

\subsection{Twitter Dataset on Russian Troll Activity}
The secondary data set used in this investigation was derived from the work of Addawood et al. \cite{addawood2019linguistic}, examining the operations of Russian trolls on Twitter around the time of the 2016 U.S. Presidential election. The dataset encompasses information from 2,752 Russian troll accounts identified and released by the U.S. Congress. Using Crimson Hexagon, a social media analytics tool, the researchers collected a significant volume of tweets and retweets (1,226,185 in total), produced by these trolls, with approximately 27\% of this content in the Russian language.
% Using Crimson Hexagon, a social media analytics tool, the researchers collected a significant volume of tweets and retweets (1,226,185 in total), produced and later erased by these trolls during the election year, with approximately 27\% of this content in the Russian language.

The dataset was further enriched by incorporating non-troll tweets, collected through two distinct methodologies. Initially, tweets were collected using a compilation of hashtags and keywords directly associated with the 2016 election. The second method involved collecting tweets from the same users while purposely omitting the aforementioned election-related terms, with a careful exclusion of any user who had retweeted troll-affiliated content. This comprehensive approach resulted in an extensive corpus of 12.36M tweets, generated by 1.17M unique user accounts.

\subsection{Dataset Generation for LLM Analysis}
For the purpose of our analysis, the datasets were selectively curated to focus on five countries: Egypt, UAE, Ecuador and Venezuela, from the Twitter Information Operations archive, and Russia, from the Twitter Dataset on Russian Troll Activity. 
We group together the accounts that Twitter associated with Egypt and the UAE influence campaigns, based on Twitter's reports \cite{gadde2020additional} and previous work \cite{wang2023evidence,ezzeddine2022characterizing} that revealed that their activity was primarily linked to an IO originating from both countries and targeting Iran and Qatar. 

The data set generation process was implemented as follows. Initially, we calculate the median time interval between the first and last tweets for each influence operation. Tweets preceding and including this median time constitute the training set, with a capping mechanism limiting each user to a maximum of 100 randomly selected tweets to curtail the dataset's volume. Tweets after the median time serve as the basis for the test set. To maintain the integrity of the data set and avoid information leakage, any users featured in the training set were omitted from the test set. The selection criteria for the test set mirrored those of the training set, with an identical cap on tweets.

% Given computational constraints, we further distill the dataset for tweet classification tasks. A random subset of tweets was selected: 25 for Venezuela and 10 each for Egypt \& UAE and Russia campaigns. 
In our study, we perform both tweet and user classification tasks. 
For the former, we further distill the dataset due to computational constraints, leveraging only a random subset of ten tweets for every user.
% : 25 each for Ecuador and Venezuela, and 10 each for Egypt \& UAE and Russia campaigns. 
For both tasks, we comprehensively include original tweets and retweets. Tables \ref{table:users} and \ref{table:tweets} offer an in-depth breakdown of the datasets, cataloging the total user counts and distinguishing between \textit{driver} of information campaigns and \textit{organic}, control accounts, in addition to detailing the aggregate tweet counts.

\begin{table}[ht]
    % \caption*{\textbf{Number of Users}}
    \centering 
    \begin{tabular}{|l|l|l|l|l|}
    \hline
    \textbf{Country} & \textbf{Train/Test} & \textbf{Driver} & \textbf{Organic} & \textbf{Driver \%}  \\
    \hline \hline
    Russia & Train & 200 & 10,000 & 2\% \\
    Russia & Test & 200 & 10,000 & 2\% \\
    Egypt\&UAE & Train & 1,000 & 4,000 & 20\% \\
    Egypt\&UAE & Test & 625 & 2,500 & 20\% \\
    Ecuador & Train & 325 & 1,625 & 20\% \\
    Ecuador & Test & 250 & 1,250 & 20\% \\
    Venezuela & Train & 500 & 2,500 & 25\% \\
    Venezuela & Test & 325 & 1,625 & 25\% \\
    \hline
    \end{tabular}
    \\[10pt]
    \caption{Number of users for each dataset. For each country, the table reports the division into training and test sets, the number of driver and organic users, and the percentage of driver users.}
    \label{table:users}
\end{table}

\begin{table}[ht]
    % \caption*{\textbf{Number of Tweets}}
    \centering 
    \begin{tabular}{|l|l|l|l|l|}
    \hline
    \textbf{Country} & \textbf{Train/Test} & \textbf{Driver} & \textbf{Organic} & \textbf{Driver \%}  \\
    \hline \hline
    Russia & Train & 1,782 & 71,922 & 2.48\% \\
    Russia & Test & 2,159 & 92,602 & 2.33\% \\
    Egypt\&UAE & Train & 8,898 & 35,983 & 24.73\% \\
    Egypt\&UAE & Test & 7,234 & 29,000 & 24.95\% \\
    Ecuador & Train & 5,276 & 32,199 & 16.39\% \\
    Ecuador & Test & 7,832 & 36,387 & 21.52\% \\
    Venezuela & Train & 10,057 & 50,949 & 19.73\% \\
    Venezuela & Test & 8,950 & 42,084 & 21.26\% \\
    \hline
    \end{tabular}
    \\[10pt]
    \caption{Number of tweets for each dataset. For each country, the table reports the division into training and test sets, the number of driver and organic tweets, and the percentage of driver tweets.}
    \label{table:tweets}
\end{table}

\section{Methods}
In a scenario where social media platforms are increasingly leveraged for influence campaigns, text, user information, and network structures are essential to detecting and understanding the dynamics of these operations. The key to effective detection and analysis of influence campaigns lies in comprehending and modeling the interconnections between these crucial components. Motivated by this requirement, our study undertakes two different tasks: \textit{(i)} tweet classification, and \textit{(ii)} user classification.
The former involves examining individual tweets to identify those that are part of an influence campaign. The latter focuses on distinguishing key actors who are actively driving these influence efforts from genuine, legitimate users. Large Language Models (LLMs) \cite{naveed2023comprehensive} are utilized for classification purposes.   The following sections provide a more detailed exploration of how our methodologies leverage this technology.

\subsection{Tweet Classification via LLMs}{\label{subsec:tweet_classification}}
For tweet classification, we envision the utility of employing LLMs to analyze and understand the structure and distinctive features of tweet messages associated with an influence campaign. In particular, the classification is carried out using the open-source LLM \textit{Llama 2} released by Facebook \cite{touvron2023llama}. The advantage of this open-source model lies in its ability to run locally, thereby eliminating the need to share data with external parties. % Furthermore, it prevents malicious actors from accessing the trained models and using them in a simulated environment to rapidly generate new techniques. The most direct interaction with the model is through zero-shot prompting, which involves providing a question or instruction. However, for tasks that require a deeper understanding of context, we can utilize few-shot prompting \cite{NEURIPS2020_1457c0d6}. This technique guides our model by offering a handful of examples, each representing a specific instance of the task. As our model uses these examples to understand the task and generate appropriate responses, the function can be restructured to incorporate these examples as part of the input.
We utilize the Microsoft Guidance  \cite{GuidanceAI2023} repository to implement both zero-shot \cite{NEURIPS2022_8bb0d291} and few-shot prompting techniques \cite{parnami2022learning}. This repository comes with a comprehensive set of tools and guidelines, specifically designed to facilitate the execution of these prompting techniques. A key feature of this repository, is its ability to enforce a binary output as a response.

As a final technique, we implemented a more sophisticated approach known as \textit{fine-tuning} \cite{DBLP:journals/corr/abs-2109-01652}, which is a training technique that learns the structure of the data, rather than solely relying on prompting. It employs a form of supervised learning where both tweets and labels are provided. For carrying out the fine-tuning process, we utilized the LLaMA Factory repository  \cite{llama-factory}.

\subsubsection{Zero-Shot Prompting}
Zero-Shot Prompting is a technique that allows our model to generate responses without having seen any examples of the task at hand \cite{NEURIPS2022_8bb0d291}. This is achieved by providing the model with a ``prompt'' in the form of a plain question or statement, which guides the model's response generation. The model then uses its understanding of language and context to generate a response, which in this case is a binary variable. 
% An example of this can be seen in the figure below, where the prompt is provided to the model and the resulting classification is generated. 
This approach is particularly useful for tasks where the context can be easily encapsulated in a single prompt, and where the model's pre-existing knowledge and understanding of language are sufficient to generate an appropriate response.

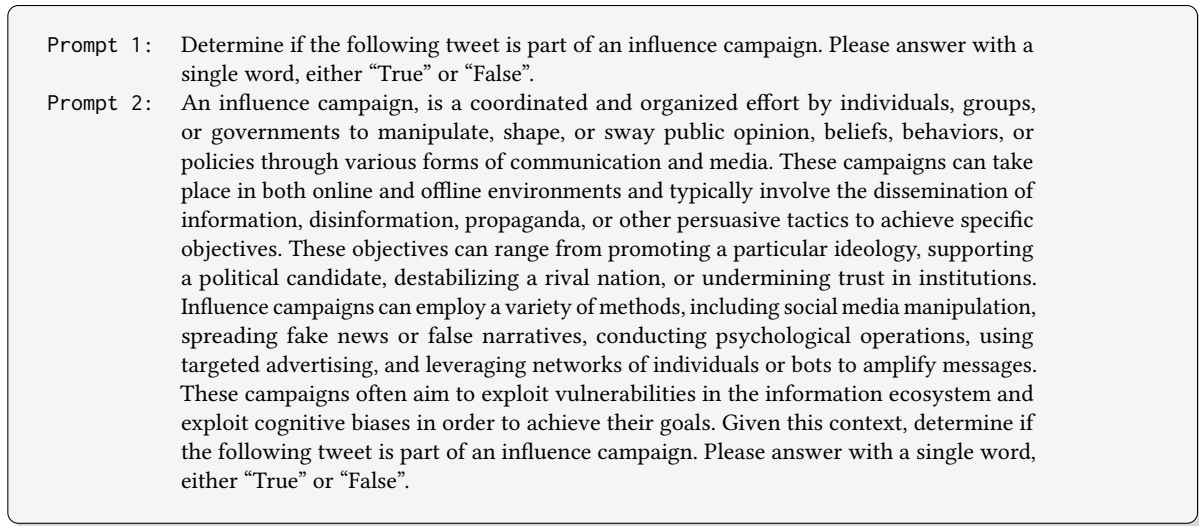
\begin{figure}[t]
    \centering
    \begin{tikzpicture}
        \node[draw, fill=mygray, rounded corners, drop shadow={fill=black!30, shadow xshift=3pt, shadow yshift=-3pt, opacity=0.5}, inner sep=10pt] {
            \begin{minipage}{1.0\textwidth}
                \begin{tabular}{l p{0.75\textwidth}}
                    \texttt{Prompt 1:} & Determine if the following tweet is part of an influence campaign. Please
answer with a single word, either ``True'' or ``False''. \\
  \texttt{Prompt 2:} & An influence campaign, is a coordinated and organized effort by individuals, groups, or
governments to manipulate, shape, or sway public opinion, beliefs, behaviors, or policies through
various forms of communication and media. These campaigns can take place in both online and
offline environments and typically involve the dissemination of information, disinformation,
propaganda, or other persuasive tactics to achieve specific objectives. These objectives can range
from promoting a particular ideology, supporting a political candidate, destabilizing a rival nation,
or undermining trust in institutions. Influence campaigns can employ a variety of methods,
including social media manipulation, spreading fake news or false narratives, conducting
psychological operations, using targeted advertising, and leveraging networks of individuals or bots
to amplify messages. These campaigns often aim to exploit vulnerabilities in the information
ecosystem and exploit cognitive biases in order to achieve their goals. Given this context,
determine if the following tweet is part of an influence campaign. Please
answer with a single word, either ``True'' or ``False''. \\
                \end{tabular}
            \end{minipage}
        };
    \end{tikzpicture}
    \caption{Prompts in the zero-shot setting related to Influence Campaigns}
     \label{fig:prompt_12}
\end{figure}

\begin{figure}[t]
    \centering
    \begin{tikzpicture}
        \node[draw, fill=mygray, rounded corners, drop shadow={fill=black!30, shadow xshift=3pt, shadow yshift=-3pt, opacity=0.5}, inner sep=10pt] {
            \begin{minipage}{1.0\textwidth}
                \begin{tabular}{l p{0.75\textwidth}}
                    \texttt{Prompt 3:} & Determine if the following tweet is part of an InfoOps campaign. Please
answer with a single word, either ``True'' or ``False''.\\
  \texttt{Prompt 4:} & InfoOps refer to a coordinated effort by individuals or groups to manipulate or shape public opinion
on Twitter by spreading false or misleading information. These operations can have various
objectives, including: - Political Manipulation: Some information operations aim to influence
political events, such as elections or policy decisions, by spreading misinformation or promoting
particular candidates or ideologies.
- Social Division: Others may seek to sow discord and amplify existing social or political divisions
by disseminating inflammatory content or exploiting sensitive issues.
- Brand or Reputation Management: Some businesses or organizations may use Twitter
information operations to manage their online reputation by spreading positive narratives or
suppressing negative ones.
- Malicious Activities: In some cases, these operations can involve cyberattacks, identity theft, or
the dissemination of harmful malware through links shared on Twitter.
- Amplification of Propaganda: State-sponsored actors or non-state actors may use Twitter to
amplify propaganda, particularly in the context of geopolitical conflicts.
Given this context, determine if the following tweet is part of an InfoOps campaign. Please
answer with a single word, either ``True'' or ``False''. \\
                \end{tabular}
            \end{minipage}
        };
    \end{tikzpicture}
    \caption{Prompts in the zero-shot setting related to InfoOps Campaigns}
    \label{fig:prompt_34}
\end{figure}
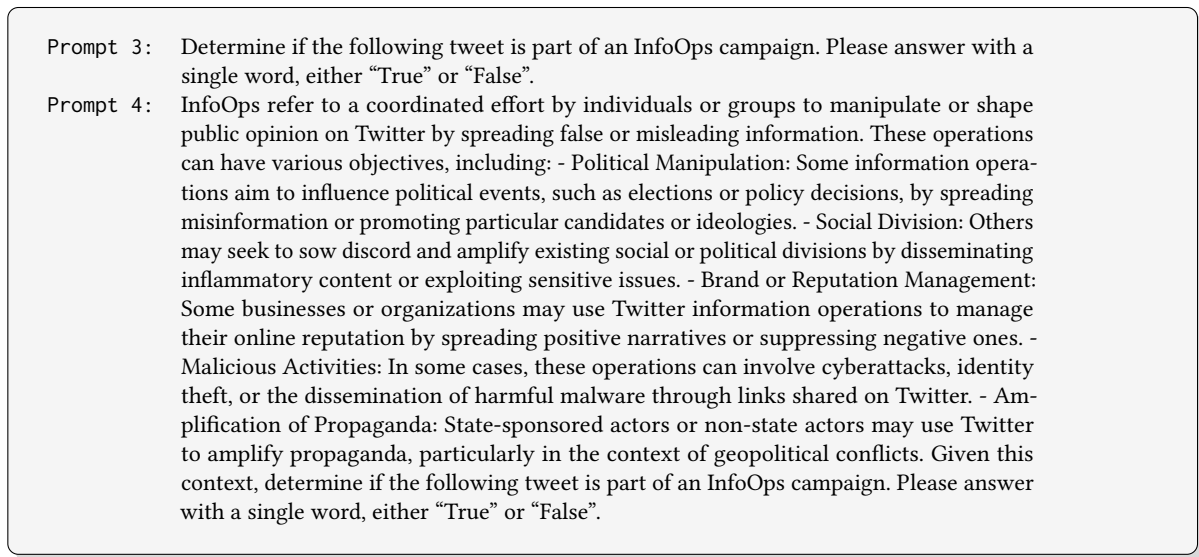

In Figure \ref{fig:prompt_12}, we display the prompts utilized in our Zero-Shot Prompting approach. These prompts specifically incorporate the term ``Influence Campaign'', presented in two variations - with and without a definition. Similarly, Figure \ref{fig:prompt_34} illustrates prompts that include the term ``InfoOps'' (abbreviation commonly used to indicate \textit{information operations}), also here showcased in two formats - with and without a definition. The definitions are sourced from ChatGPT \cite{openai_2023}.

% \begin{figure}[h]
%     \centering
%     \includegraphics[width=0.45\textwidth]{prompt12.png}
%     \caption{Influence Campaigns prompts}
%     \label{fig:prompt_12}
% \end{figure}

% \begin{figure}[h]
%     \centering
%     \includegraphics[width=0.45\textwidth]{prompt34.png}
%     \caption{InfOps prompts}
%     \label{fig:prompt_34}
% \end{figure}

\subsubsection{Few Shots Prompting} 
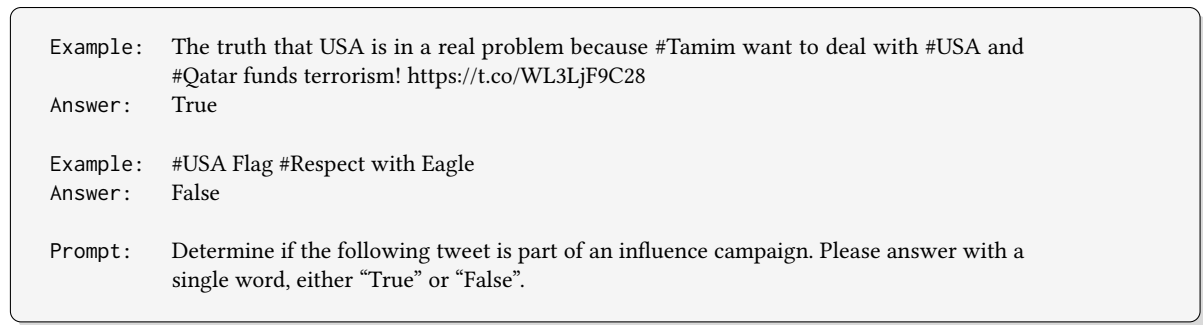
\begin{figure}[t]
    \centering
    \begin{tikzpicture}
        \node[draw, fill=mygray, rounded corners, drop shadow={fill=black!30, shadow xshift=3pt, shadow yshift=-3pt, opacity=0.5}, inner sep=10pt] {
            \begin{minipage}{1.0\textwidth}
                \begin{tabular}{l p{0.75\textwidth}}
                    \texttt{Example:} & 
                       The truth that USA is in a real problem because \#Tamim want to deal with \#USA and \#Qatar funds terrorism! https://t.co/WL3LjF9C28\\
                      
                    \texttt{Answer:} & True \\
                    \\
                     \texttt{Example:} & \#USA Flag \#Respect with Eagle\\
                    \texttt{Answer:} & False \\
                    \\
                      \texttt{Prompt:} & Determine if the following tweet is part of an influence
campaign. Please
answer with a single word, either ``True” or ``False''.\\
                    %   \\
                    % \texttt{Tweet:} & We’ve been there with \texttt{@}RoarLoudTravel!! Love the message!!! \#StreetArt \#StreetArtChat \#MondayMotivation \#Travel https://t.co/PTmEfAphnq\\
                \end{tabular}
            \end{minipage}
        };
    \end{tikzpicture}
    \caption{Example of prompt in the few-shots setting}
    \label{fig:few_shots}
\end{figure}

According to the principles of few-shot prompting \cite{parnami2022learning}, we can adjust our model to incorporate some labeled examples. By doing so, we aim to enhance the model's ability to generalize from these examples to unseen instances of the task. Specifically, we form our prompts to include two examples of tweets for classification: one from an organic user and one from an information operation driver. In doing so, we aim to provide our model with a clearer understanding of the task and improve its performance in detecting influence campaign messages. An example of prompt is illustrated in Figure \ref{fig:few_shots}.
% Additional examples are available in the supplementary material.
% \begin{figure}[h]
%     \centering
%     \includegraphics[width=0.45\textwidth]{few_shots_examples.png}
%     \caption{Few Shots Example}
%     \label{fig:few_shots}
% \end{figure}

\subsubsection{Fine-tuning}
Fine-tuning is a key step in adapting \textit{LLaMa 2} to the specific task of detecting influence campaign messages on Twitter. This process involves tuning the model to recognize the specific patterns associated with messages belonging to influence campaigns or grassroots conversations. Notably, this process involves adjusting only a subset of the model's weights to better fit our task, rather than retraining all the weights, which could be both computationally and economically daunting \cite{DBLP:journals/corr/abs-2004-08900}.

We create JSON files for both the training and testing phases, specifically structured to fine-tune the model on tweet textual content. Every JSON file encapsulates the essential information needed for fine-tuning, including the input (tweet text), the instruction (prompt), and the label. We then employ the repository \textit{LLaMa Factory} \cite{llama-factory}, to fine-tune the model and adjust the weights to our specific task. Upon completion of the fine-tuning process, the adapted model can then be employed to predict labels, enabling us to classify unseen tweets and identify potential influence campaigns.

\subsection{LLM-empowered User Classification}
To perform user classification, we use a modified version of \textit{LLaMa 2}, specifically LLaMA-2-7B-32K, developed by Together Computer \cite{togethercomputer2023llama2}. This enhanced model is designed to accommodate a larger input space of 32,000 tokens, surpassing the standard model's maximum context limit of 8,000 tokens. This key enhancement is instrumental in managing the extensive information associated with user classification.

We fine-tune multiple models using the LLaMa Factory by incorporating various types of information. The fine-tuned models are then used for classification purposes. We propose four different models, and their combination in a unified model, leveraging users' \textit{interactions}, \textit{centralities} in the interaction network, \textit{metadata}, and shared \textit{content}.
In the next sections, we provide a detailed description of this process, discussing the diverse types of information used for fine-tuning, and the application of the models for classification. 

\subsubsection{Interaction-based User Classification}
\label{rt_net}
The core idea of this model is to extract the structure of the interaction network based on the retweets exchanged between Twitter users. As \textit{LLaMa 2} models the relationships in text, it can be used to represent the network once its description is converted into words. This allows us to directly use the output of the model, bypassing the complexity of traditional network extraction methods. In order to transcribe the network into text, we followed the paradigm developed by Ye et al.  \cite{ye2023natural}.

We construct a network of retweets to capture the interactions between users. In this network, each user is represented as a node, and an edge represents a retweet from the user who retweets to the author of the original tweet. 
%The process of extracting the network can be described as follows. Initially, we identify all users and their retweets based on the tweets and their associated metadata. Subsequently, these retweets are utilized to form edges between users in our network. 
The training retweet network is created using all the tweets from the training set. In this network, user \textit{i} is connected to user \textit{j} if user \textit{i} has retweeted at least once user \textit{j}. Similarly, the test retweet network is constructed using all retweets at our disposal. It is important to note that users in the training set do not overlap with users in the test set.

Given this formulation, we generate JSON files with specific instructions and inputs. The instruction is to ``Determine if the user is actively driving an influence campaign.'' The input is structured as ``User \textit{i} is connected to \textit{j, k, m}, ...'' and the output is a binary variable (True vs. False) indicating whether the user is driving an influence campaign. Accordingly, we fine-tune our model and classify accounts based on the model outputs.

\subsubsection{Centrality-based User Classification}
Centralities measures are a crucial element in identifying coordinated actions \cite{luceri2023unmasking}. In this work, we employ different properties of network centrality \cite{saxena2020centrality} to identify drivers of influence campaigns. In particular, we compute the degree, eigenvector, and PageRank centralities based on the retweet network (cf. \S\ref{rt_net}). These metrics are used to generate JSON files for training and testing. Each JSON file contains the following information: The instruction ``Determine if the user is actively driving an influence campaign''; the input ``User \textit{i} has a degree centrality of \textit{x}, an eigenvector centrality of \textit{y}, and a PageRank centrality of \textit{z}''; and the output is either ``True'' or ``False''. These JSON files are used for fine-tuning and classification. 

\subsubsection{Metadata-based User Classification}
Metadata can also play a crucial role in enhancing the performance of our model \cite{addawood2019linguistic}. Specifically, for each tweet, we extract metadata that provides additional context about the tweet,
% . All the generated metadata reflects the characteristics of the tweet and its author, providing valuable information that 
which can help our model better understand the context in which the user operates. 
% The formal process of metadata extraction can be described as follows: for each user, we extract their metadata, concatenate it, and add it to the query. 
The analyzed metadata includes hashtags, URLs, and user mentions derived from the user's tweets. The processing of this metadata involves the removal of duplicates. Specific modifications are considered for URLs. For Twitter and Telegram URLs, the process involves extracting the last segment of the link. For example, ‘https://t.me/1234’ becomes ‘1234’ and ‘https://t.co/5678’ becomes ‘5678’. For all other URLs, we consider only their domain, e.g., ‘www.foxnews.com/politics’ becomes ‘www.foxnews.com’. 

    The construction of the JSON files is similar to the models previously discussed: The instruction is ``Determine whether the user is actively driving an influence campaign''; The input is ``The user has the following metadata: hashtags, URLs, and mentions''. The expected output is a binary variable that indicates whether the user is driving an influence campaign. As before, we use this framework to fine-tune the model and perform our classification task.

\subsubsection{Content-based User Classification}
 This model leverages the \textit{tweet classification} task to generate a score to classify users as either drivers of influence campaigns or organic users.
This process involves multiple steps, starting from the division of training tweets into train and validation sets, where 90\% is assigned to the train set and 10\% to the validation set. JSON files are then generated for the tweets, with the aim of determining if a given tweet is part of an influence campaign. The input of the model is the text of the tweet, and the output is a binary variable indicating whether the tweet belongs to an influence campaign.

Fine-tuning is performed on the tweets in the train set, and predictions are made on the validation and test set. The user score is computed by taking the average of test outcomes (0 for \textit{False} and 1 for \textit{True}) of user's classified tweets. The score for control users is expected to be near zero, while it should be close to one for driver users. The threshold for determining if the user is driving an influence campaign is tuned by maximizing the AUC on the validation set.
The classification on the test set uses the threshold established in the validation step to perform user classification.

\subsubsection{Multi-Input Model}
\label{sec:multi-input}
Finally, we develop a model that combines different input sources. This approach aims to leverage the strengths of each individual data source, potentially enhancing the overall predictive power of our model. As in the models described above, the instruction in the JSON files aims to determine if a user is driving an influence campaign. As input, we use a combination of the inputs utilized in the other models, i.e., retweet connections, centralities, metadata, and content-based score. This diverse suite of features is concatenated, forming a multi-input model that is fine-tuned to perform user classification.
% This allowed our model to learn from a diverse set of features.
% Once the JSON files were generated and the inputs concatenated, the next step was to fine-tune the model and make predictions.

%%RESULTS

\begin{table*}[t!]
  \centering
  \begin{threeparttable}
  % \caption*{\textbf{Tweet Classification Results}}
  \begin{tabular}{lcccc}
    \toprule
    {Test} & {Precision (\%)} & {Recall (\%)} & {F1-Score (\%)} & {AUC (\%)} \\
    \midrule
    Zero-Shot Prompt 1 & 16.56 $\pm$9.93 &  55.80 $\pm$10.99 &  22.95 $\pm$12.02 &  49.16 $\pm$7.07  \\
    Zero-Shot Prompt 2 &  14.21 $\pm$7.19 &   95.61 $\pm$5.03 & 23.94 $\pm$11.69 &   49.66 $\pm$0.54 \\
    Zero-Shot Prompt 3 & 13.56 $\pm$7.36 &   35.92 $\pm$9.01 & 18.14 $\pm$9.69 &   46.79 $\pm$4.41 \\
    Zero-Shot Prompt 4 & 9.90 $\pm$7.89 &   \textbf{98.84 $\pm$0.42} & 17.12 $\pm$13.10 &   49.90 $\pm$0.18 \\
    Few-Shot & 0.00 $\pm$0.00 &    0.00 $\pm$0.00 &  0.00 $\pm$0.00 &   49.99 $\pm$0.01  \\
    Fine-Tuning & 64.74 $\pm$15.56 &  46.93 $\pm$28.24 & 46.95 $\pm$14.52 &   \textbf{69.61 $\pm$9.68} \\
Linguistic Cues \cite{addawood2019linguistic} & \textbf{91.20 $\pm$6.47} &   97.41 $\pm$2.98 & \textbf{94.08 $\pm$3.90} &  69.23 $\pm$12.73 \\
    \bottomrule
  \end{tabular}
  \caption{Comparative classification performance of the tweet classification task for different methods across four influence campaigns.}
\label{table:tweets_classification_table}
  \end{threeparttable}
\end{table*}

% EF: DO NOT MOVE THESE
\begin{figure*}[t!]
    \centering
    \includegraphics[width=1.0\textwidth]{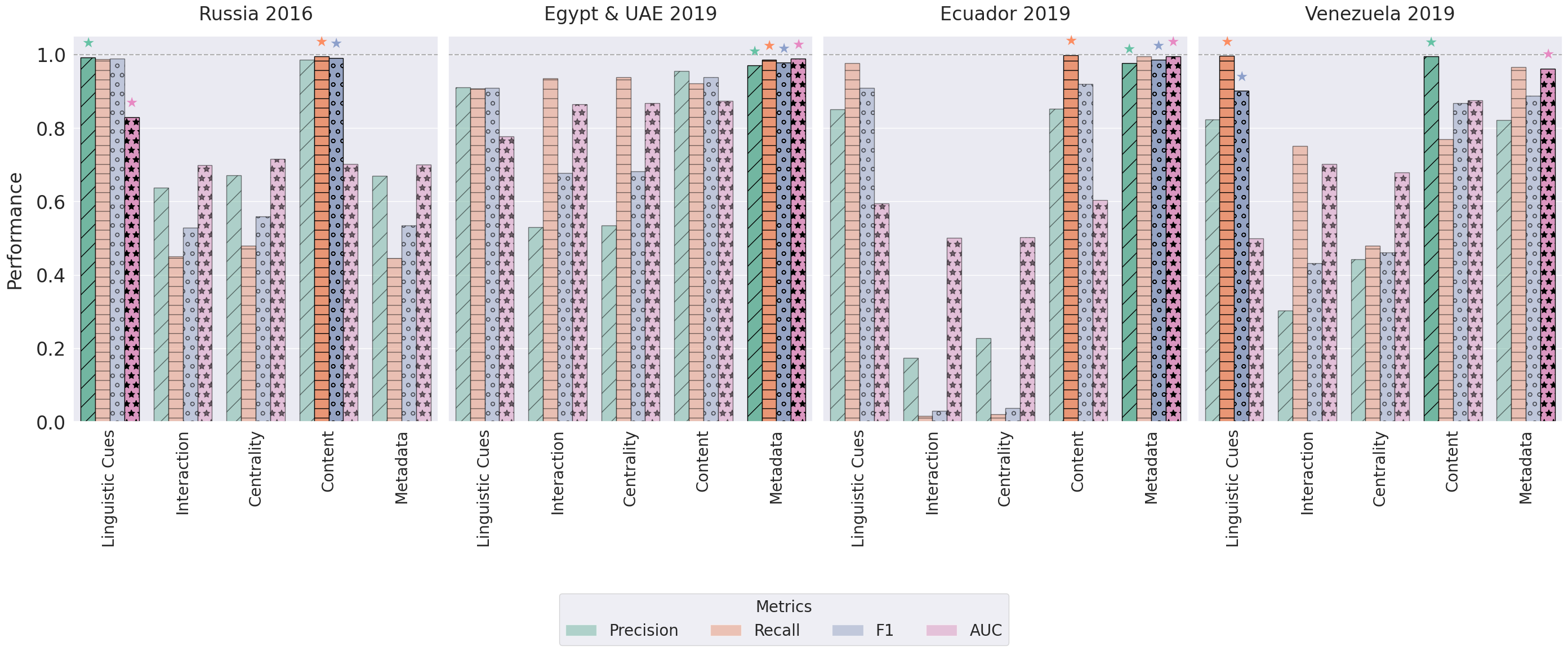}
    \caption{Comparative analysis of five methods across four influence campaigns. The  best-performing technique is highlighted.}
    \label{fig:user_classification_plot}
\end{figure*}

\section{Results}
In our empirical analysis, we engage in two main tasks aimed at uncovering the messages and actors driving influence campaigns. The first task is tweet-centric and aims at classifying content as belonging or not to influence operations. The second task focuses on users, assessing their involvement in such campaigns by examining their tweets, metadata, connections, and position in the interaction network. We use the linguistic approach proposed by Addawood et al.  \cite{addawood2019linguistic} as our baseline model and benchmark this technique against our proposed models. Our investigation spanned four distinct influence campaigns, as described in Section \ref{subsec:tweet_classification}.

\subsection{Tweet Classification}

For the tweet classification task, we compare our proposed zero-, few-shot, and fine-tuned models with a linguistic approach \cite{addawood2019linguistic}.
The results of the tweet classification task are presented in Table \ref{table:tweets_classification_table} by averaging the classification performance across the four datasets.

The linguistic approach exhibits superior performance in terms of precision and F1-score compared to our proposed methods. The results concerning zero- and few-shot methods demonstrate the considerable challenge of this task when limited or no information is provided for training a model. Interestingly, when definitions are supplied (Prompt 2 and Prompt 4) in a zero-shot setting, our models attain near-perfect recall. However, this achievement is coupled with limited precision.
Additionally, it is important to emphasize the challenges encountered in the few-shot setting. When examples of driver and organic tweets are given, the model often predicts false outcomes almost consistently. This results in precision, recall, and F1 scores dropping to zero, given the minimal or nonexistent count of true positives. Nevertheless, the subpar classification performance could stem from the manual selection of tweets provided to the LLM. This process may either lack representativeness of influence messages or introduce inherent biases. Future efforts will be directed towards addressing this issue.

Finally, when assessing the AUC, our fine-tuned based on Llama Factory allows us to achieve superior results compared to the baseline technique, highlighting its potential for the user classification task, and in particular for the \textit{content}-based model.

% indicating a higher true positive rate when considering false positives.

\begin{table*}[t!]
  \centering
  \begin{threeparttable}
  % \caption*{\textbf{User Classification Results Table}}
  \begin{tabular}{lcccc}
    \toprule
    {Test} & {Precision (\%)} & {Recall (\%)} & {F1-Score (\%)} & {AUC (\%)} \\
    \midrule
Linguistic Cues \cite{addawood2019linguistic} &  89.43 $\pm$6.71 &   \textbf{96.71 $\pm$3.60} &   92.76 $\pm$3.73 &  67.55 $\pm$13.81 \\

Centrality &  46.92 $\pm$16.69 &  47.98 $\pm$33.56 &  43.48 $\pm$25.07 &   69.17 $\pm$13.39 \\

Content & \textbf{94.74 $\pm$5.84} &   92.12 $\pm$9.61 &   \textbf{92.92 $\pm$4.55} &  76.37 $\pm$12.04 \\

Interaction  & 41.14 $\pm$18.89 &  53.82 $\pm$35.92 &  41.65 $\pm$24.79 & 69.18 $\pm$13.33 \\

Metadata  &    85.94 $\pm$13.02 &  84.82 $\pm$24.07 &  84.67 $\pm$19.03 & \textbf{91.18 $\pm$12.67} \\

    \bottomrule
  \end{tabular}
  \caption{Comparative classification performance of the user classification task for different methods across four influence campaigns.}
\label{table:users_classification_table}
  \end{threeparttable}
\end{table*}

\subsection{User Classification}{\label{subsec:user_classification}}
We conducted a series of experiments to assess the performance of our proposed methods versus the baseline approach \cite{addawood2019linguistic}. 
The results of these experiments are illustrated in Figure \ref{fig:user_classification_plot}, whereas Table \ref{table:users_classification_table} aggregates the results of the different campaigns.

There are a few noteworthy observations. First, among the proposed models, the content-based approach consistently demonstrates strong classification accuracy across various campaigns, either outperforming the state-of-the-art approach based on linguistic cues \cite{addawood2019linguistic} or delivering comparable results. Second, the metadata-based model also achieves promising classification results, with the highest average AUC among the evaluated approaches. Other models, such as those based on interactions and centralities, show limited classification capabilities across campaigns.
Finally, the baseline approach excels in identifying the 2016 Russian campaign but experiences diminishing efficacy with more recent influence operations, such as those from Egypt \& UAE, Venezuela, and Ecuador.
Overall, these results indicate the potential of LLMs to adapt to more sophisticated and unseen campaigns in which malicious actors could evade existing detection solutions.

% As a result, we advocate for a two-step strategy to achieve optimal performance: initially, employ methods with human-in-the-loop detection to identify new users involved in influence operations (infops) and control users; subsequently, apply our methods to these newly identified users.

\paragraph{Ablation Study}
We conduct an ablation study to evaluate the effectiveness of different types of information and identify the most relevant feature. In each ablation test, we use models based on retweet interactions, centralities, content, metadata, and their combinations, as described in Section \ref{sec:multi-input}. 

\begin{figure}[t!]
    \centering
    \includegraphics[width=0.67\textwidth]{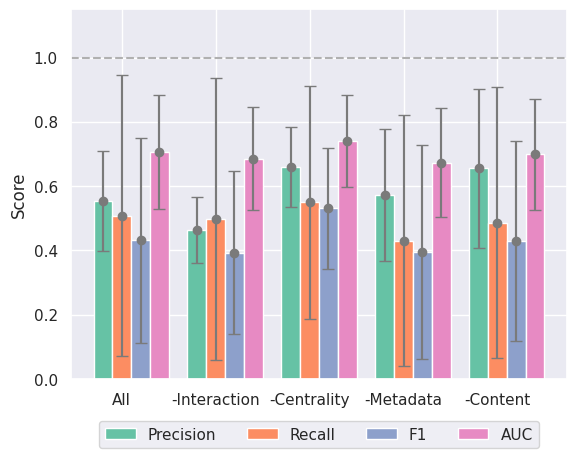}
    \caption{Ablation study: performance of models using all feature dimensions, and respectively after removing \textit{interaction}, \textit{centrality}, \textit{metadata}, or \textit{content} features.}
    \label{fig:ablation}
\end{figure}

The results of these tests are shown in Figure \ref{fig:ablation}. A test marked as ``-X'' indicates that we use a model with all features except ``X''.
Notably, classification performance remains consistent when we employ all features or exclude just one. However, models based on a single feature at a time generally achieve superior performance (\textit{cf.} Fig. \ref{fig:user_classification_plot}). This prompts questions about potential interferences between these combinations and underscores the need for further research to enhance the adaptability of LLMs to diverse inputs.
% longer contexts with various types of inputs.

% This suggests that while the integration of diverse features might represent a promising strategy, it appears that a  at a time can lead to improved results. 

\section{Conclusions}
In this study, we introduced novel methodologies to identify influence campaigns on social media building upon the general-purpose language modeling strengths of LLMs. We found that while current state-of-the-art methods are effective, their performance diminishes when applied to new influence campaigns, highlighting the need for innovative dynamic detection methods.
Our methodology addresses this shortfall by incorporating elements such as content, network structures, and user metadata. These elements are subsequently transformed into a textual format compatible with Large Language Models (LLMs).
Empirical results indicate that our model not only successfully identifies influence campaigns but also adapts to the multilingual and multifaceted dimensions of social media, thereby enhancing the robustness of influence detection mechanisms.

Despite the encouraging results, we recognize the need for continuous improvement. Future research should focus on enhancing the flexibility of LLMs to accommodate multiple input data sets to improve overall results.  Furthermore, the potential to broaden the scope of this research utilizing different and larger models, such as those with 40B and 70B parameters, underscores an exciting avenue for future exploration. This presents a compelling trajectory for the advancement of this field.

\paragraph{Limitations.}
This research is not exempt from limitations. First, we utilize only a subset of influence campaigns released in the Twitter information operations archive \cite{gadde2020additional}. This limitation is due to constraints related to computational resources and the computing time required for our experiments. Nevertheless, the diversity of the analyzed campaigns ensures the robustness of our results.

Second, for the same resource-related reasons, the models developed for both the text and user classification tasks rely on only a subset of available tweets. The presented performance may, therefore, potentially improve when employing a larger set of tweets.

Third, we adopted a manual approach for selecting few-shot examples. This involved reviewing and choosing tweets that best represented our two classes of accounts. However, we recognize that this selection method may introduce biases. The selection of tweets is subjective and may not precisely capture the distinction between messages generated by drivers and organic users. These factors could have impacted the results associated with the few-shot approach, demonstrating poor classification performance. Future work will explore alternative methods for selecting few-shot examples to address these issues.

Finally, our identification of influence campaign drivers relies on users recognized by Twitter, and the precise methods behind this identification remain undisclosed. Potential biases in data collection and the potential misclassification of accounts can affect our models' detection effectiveness.

% the sharing activities, messages, and metadata of organic users may vary in frequency compared to those of the drivers of influence campaigns. These distinctions suggest that control users may inherently form distinct networks, not solely due to their non-IO status.

\paragraph{Ethical considerations.}
Prioritizing user privacy, we ensured that all control data were anonymized before any analysis was performed. It is essential to recognize that, despite the rigor of our approach, there is a possibility that our model may inadvertently misidentify legitimate user accounts as those involved in influence campaigns. This requires a careful and critical analysis of the results obtained. Furthermore, it is conceivable that the actors behind influence campaigns could be mistakenly classified as regular control accounts, which could allow the continued propagation of misinformation or fraudulent activities. Therefore, we recommend using our model within a broader toolkit designed to incorporate other behavioral, textual, and metadata features to improve the classification accuracy between authentic users and those engaged in influence operations.

\begin{acks}
Work supported in part by DARPA (contract \#HR001121C0169).
\end{acks}

% \newpage
\bibliographystyle{ACM-Reference-Format}
\balance
\bibliography{sample-base}
% \appendix









\end{document}